\documentclass[12pt]{article}
\usepackage{latexsym}
\usepackage{amstext}
\usepackage{amsmath}
\usepackage[normalem]{ulem}
\usepackage{subfigure}
\usepackage{epsfig, graphicx}
\usepackage{color}
\usepackage{amsfonts,bm,color}
\usepackage{verbatim}
\usepackage{float}
\usepackage{latexsym}
\usepackage{stmaryrd}
\usepackage{setspace}

\newcommand{\ds}{\displaystyle}
\newcommand{\beq}{\begin{eqnarray}}
\newcommand{\eeq}{\end{eqnarray}}
\newcommand{\beqq}{\begin{eqnarray*}}
\newcommand{\eeqq}{\end{eqnarray*}}
\newcommand{\p}{\partial}

\newcommand{\eps}{\varepsilon}

\newcommand{\x}{\mbox{\boldmath$x$}}

\newcommand{\s}{\mbox{\boldmath$s$}}

\newcommand{\SC}{\textsc}

\begin{document}
  \title{Voltage laws for three-dimensional microdomains with cusp-shaped funnels derived from Poisson-Nernst-Planck equations}
  \author{J. Cartailler\footnote{ Ecole Normale Sup\'erieure, 46 rue d'Ulm 75005 Paris, France.}
   \,and D. Holcman$^*$\footnote{Mathematical Institute, University of Oxford, Andrew Wiles Building, Woodstock Rd, Oxford OX2 6GG, United Kingdom. \newline {{\bf Corresponding author email:} david.holcman@ens.fr}}}
  \date{\today}

\maketitle

\begin{abstract}
We study the electro-diffusion properties of a domain containing a cusp-shaped structure in three dimensions when one ionic specie is dominant.  The mathematical problem consists in solving the steady-state Poisson-Nernst-Planck (PNP) equation with an integral constraint for the number of charges. A non-homogeneous Neumann boundary condition is imposed on the boundary.  We construct an asymptotic approximation for certain singular limits that agree with numerical simulations. Finally, we analyse the consequences of non-homogeneous surface charge density. We conclude that the geometry of cusp-shaped domains influences the voltage profile, specifically inside the cusp structure. The main results are summarized in the form of new three-dimensional electrostatic laws for non-electroneutral electrolytes. We discuss applications to dendritic spines in neuroscience.
\end{abstract}	
  %
\section{Introduction}
We study here the Poisson-Nernst-Planck (PNP) equations in three dimensional for domains containing a cups-shaped funnel. These equations are used to describe electro-diffusion processes in ionic channels \cite{Nadler,Singer} and also in neurobiological microdomains \cite{Hille,Bezanilla}, where charges are coupled though the electric field. We consider here a generic domain formed of a ball with an attached cusp-shaped funnel on its boundary. Such geometry is common in cellular neurobiology, for instance dentritic spines \cite{Harris}, where the structure cannot be reduced to 1D geometry \cite{HY2015}.  Phenomenological descriptions of electro-diffusion, using the linear cable theory, $RC$-electric circuit representation, and even electronic devices, are insufficient to describe non-cylindrical geometry \cite{Hille,HY2015}, since they assume a simple reduced one-dimensional or an overly simplified geometry.

We present here nove results about the voltage landscape based on the electro-diffusion model in various microdomains, when the condition of electro-neutrality is not satisfied and one ionic specie is dominant. The boundary is impermeable to particles (ions) and the electric field satisfies the compatibility condition resulting from Poisson's equation. Under the non-electro-neutrality assumption and with charge distributed in bounded domains, confined by a dielectric membrane, Debye's law of charge screening decaying exponentially away from a charge \cite{Debye} does not apply and long-range correlation are expected, leading to a gradient of charges in a domain with no inward current. We derived a new capacitance law for an electrolyte ball \cite{PhysD2016} and for a two-dimensional cusp \cite{NonLin2017}, where the difference of potential $V(C)-V(S)$ between the center $C$ and the surface $S$ increases, first linearly and then logarithmically when the total number of charges in the ball increases.

Our aim here is to estimate the effect of boundary curvature on three-dimensional electrical domains such as dendritic spines. In particular, we explore the effect of boundary curvature on the charge and field distribution at steady state. The curvature of neuronal dendrites and axons membranes possesses many local maxima that can modulate the channel's local electric potential \cite{Bezanilla,YusteBook,BioRXiv}. In this article, we study the effects of local curvature on the distribution of charges with no electro-neutrality. The effect of non-electro-neutrality was recently studied in \cite{PhysD2016,NonLin2017} and a long-range electrostatic length, much longer than the Debye length was found. This effect is due to the combined effects of non-electro-neutrality and di-electric boundary, which lead to charge accumulation. The cusp-shaped funnel was studied in the context of diffusion in \cite{HS2012}, but we focus here on a three-dimensional nonlinear problem with non-homogeneous Neumann boundary condition and we further extend the matched asymptotic analysis based on conformal mapping, different from the classical matched asymptotic methods \cite{Ward0,Ward4,Ward2,Ward3,Ward1}.

\begin{figure}[H]
	\center
	\includegraphics[scale=0.75]{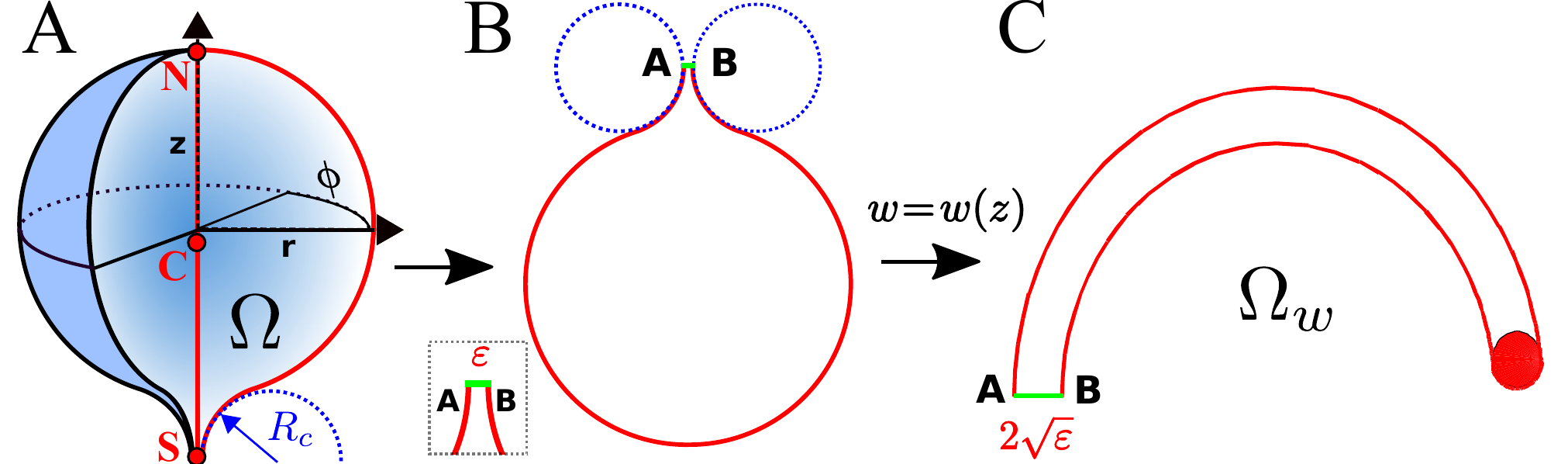}
	\caption{ {\small  {\bf Ball with a cusp-shaped funnel and image $\Omega_w$ of the domain $\Omega$ cross-section under the conformal mapping \eqref{Mobius}}
	{\bf A.} schematic representation of the domain $\Omega$, with the funnel curvature radius $R_C$, the north pole $N$, the funnel tip $S$, and the center of mass $C$.
	{\bf B-C} The neck ({\bf B}) is mapped onto the
	semi-annulus enclosed between the like-style arcs and the large disk in $\Omega$ is mapped
	onto the small red disk. The short green segment $AB$ (left) (of length $\eps$) is
	mapped onto the thick green segment $AB$ (of length $2\sqrt{\eps}+O(\eps)$).} \label{f:conf}}
\end{figure}

The manuscript is composed of three parts: in sections 1 and 2, we extend the results we have obtained in \cite{NonLin2017}, that describe the voltage in a planar cusp with homogeneous surface charge density. We then focus on an uncharged cusp for a 3D cusp-shaped funnel. In the third section, we extend the results derived in section 1 to a non-homogeneous surface charge density. We summarized now the new electrostatic laws we derived here for the difference of potential $V(C)-V(S)$ where $C$ is the center of mass of the domain and $S$ is located at the bottom of funnel (Fig. \ref{f:conf}A).

For a constant surface charge density (section \ref{s:2}, eq. \eqref{unif_diff}), the voltage difference is given by
\beq
V(C)-V(S)&=&\ds\frac{k T}{e}\left (\ln\sin^2\frac{\pi |\p\tilde\Omega|}{(e^2/kT) N\tilde\eps}-\ln\frac{2\,(e^2/kT)^2\pi^2
N^2 R_c^2 }{\left(4|\p\tilde\Omega| +(e^2/kT) N\tilde \eps\right)^2}+O(1) \right)\nonumber,
\eeq
that depends the number $N$ of ions enclosed in the domain $\tilde\Omega$, the thermal energy $kT$ and the elementary charge $e$ of the electron ($1.602\cdot10^{-19}C)$, the cusp-shaped funnel width at the base $\tilde \eps$, and its curvature radius $R_c$ (see. Fig. \ref{f:conf}A).

When the surface of the cusp does not carry any charges, the voltage difference (section \ref{s:non_uniform}, eq. \eqref{Diff_pot_uncharged}) is
\beq
V(C)-V(S)&=&\ds \frac{k T}{e}\left (- \ln  \frac{8 R_c\tilde\eps}{ \pi^4|\p\tilde\Omega_{\eps}|\left(1 +  N_{bulk}/N_{\eps}\right)}    -\ln\sin^2 \ds \frac{2|\p\tilde \Omega_\eps|}{\ds (e^2/kT) N_\eps\sqrt{ R_c \tilde\eps}} +O(1) \right),  \nonumber
\eeq
which depends on the surface $|\p\Omega_\eps|$ at the end of the funnel, the number of charges $N_{bulk}$ and $N_{\eps}$ in bulk and at the end of the funnel respectively.  When the surface charge density is non-homogeneously distributed, the potential differences (section \ref{s:sigma_diff}, formula \eqref{Delta_diff_ub_large}) are given by
\beq
V(C)-V(S)&=&\ds\frac{k T}{e}\left(\ln\sin^2\frac{\pi |\p\tilde\Omega_\eps|}{(e^2/kT) N_\eps\tilde\eps}-\ln\frac{2\,(e^2/kT)^2\pi^2
	N_{cusp}^2 R_c^2 }{\left(4|\p\tilde\Omega_{cusp}| +(e^2/kT) N_{cusp}\tilde \eps\right)^2}+O(1)  \right)\nonumber,
\eeq
which depends on the total surface charge density $N_{cusp}$ on the cusp. \\
These new electrostatics expressions are asymptotic formula derived in the limit $\tilde \eps\ll1$ and for a large number of charges. There are the main results of the present study.
\section{The Poisson-Nernst-Planck equations}\label{s:2}
The Poisson-Nernst-Planck equations is a classical model of electro-diffusion. In a domain $\tilde\Omega$, the total charge in $\tilde\Omega$ results from the sum of the positive $N_p$ and negative $N_m$ charges. The concentration of mobile ions \cite{Hille} shows an imbalance of positive negative ions $N_p\gg N_m$, such that the charges in $\tilde\Omega$ can be approximated \cite{NonLin2017} by $N$ identical positive ions with an initial density $q(\tilde\x)$ in $\tilde\Omega$.  The valence is $z$ and the total number of particles is
\begin{align}
\int\limits_{\tilde\Omega} q( \s)\,d \s=N.
\end{align}
The total charge in the domain $\tilde\Omega$ is
\beqq
Q=zeN,
\eeqq
where $e$ is the electron charge. The charge density $\rho(\tilde\x,t)$ is the solution of the Nernst-Planck equation
\begin{align}
D\left[\Delta \rho(\tilde\x,t) +\frac{ze}{kT} \nabla \left(\rho(\tilde\x,t) \nabla \phi(\tilde\x,t)\right)\right]=&\,
\frac{\p\rho(\tilde\x,t)}{\p t}\hspace{0.5em}\mbox{for}\ \tilde\x\in\tilde\Omega\label{NPE}\\
D\left[\frac{\p\rho(\tilde\x,t)}{\p n}+\frac{ze}{kT}\rho(\tilde\x,t)\frac{\p\phi(\tilde\x,t)}{\p
	n}\right]=&\,0\hspace{0.5em}\mbox{for}\ \tilde\x\in\p\tilde\Omega \label{noflux}\\
\rho(\tilde\x,0)=&\,q(\tilde\x)\hspace{0.5em}\mbox{for}\ \tilde\x\in\tilde\Omega,\label{IC}
\end{align}
where $kT$ represents the thermal energy.  The electric potential $\phi(\tilde\x,t)$ in $\tilde\Omega$ is the solution of the Poisson equation
\begin{align}
\label{poisson} \Delta \phi(\tilde\x,t) =&\,
-\frac{ze\rho(\tilde\x,t)}{\eps_r\eps_0}\hspace{0.5em}\mbox{for}\ \tilde\x\in\tilde\Omega\\
\frac{\p\phi(\tilde\x,t)}{\p
	n}=&\,-\tilde\sigma(\tilde\x,t)\hspace{0.5em}\mbox{for}\ \tilde\x\in{\p\tilde\Omega},\label{Boundary_Phi} 
\end{align}
where $\eps_r\eps_0$ is the permitivity of the medium and $\tilde\sigma(\tilde\x,t)$ is the surface charge density on the boundary $\p\tilde\Omega$.
\subsection{Steady solution in a three-dimensional ball with a cusp-shaped funnel}\label{ss:SSS}
To study the effect of a narrow funnel attached to a sphere filled with an electrolyte as illustrated Fig. \ref{f:conf}A, we study the solution of the steady-state equation \eqref{NPE}
\begin{align}
\rho(\tilde\x)=N\frac{\exp\left\{-\ds\frac{ze\phi(\tilde\x)}{kT}\right\}}
{\ds\int_{\tilde\Omega}\exp\left\{-\ds\frac{ze\phi(\s)}{kT}\right\}\,d \s},\label{N}
\end{align}
hence \eqref{poisson} results in the Poisson equation
\begin{align}
\Delta\phi(\tilde\x)=-\frac{zeN\exp\left\{-\ds\frac{ze\phi(\tilde\x)}{kT}\right\}}{\eps_r\eps_0{\ds\int_{\tilde\Omega}
		\exp\left\{-\ds\frac{ze\phi(\s)}{kT}\right\}\,d\s}}{.}\label{Deltaphi}
\end{align}
and \eqref{Boundary_Phi} gives the boundary condition
\beq\label{compatibility}
\frac{\p\phi({\tilde\x})}{\p n}=-\frac{Q}{\eps_r\eps_0 |\p \tilde\Omega|}\,\mbox{  for  }\, \tilde\x\in \p\tilde \Omega.
\eeq
Equation \eqref{compatibility} represents the compatibility condition obtained by integrating the Poisson's equation \eqref{poisson} over the domain $\tilde\Omega$, assuming the surface charge density is constant. Using non-dimensional variables, we define the normalized field
\beq \label{conversion}
\bar u(\tilde\x)=\ds \frac{ze \phi({\tilde\x})}{kT},\quad\lambda= \frac{(ze)^2N}{\eps_r\eps_0 kT},
\eeq
where $\lambda$ generalizes the Bjerrum length $l_B=e^2/kT$. The Poisson's equation \eqref{Deltaphi} reduces to
\begin{align}
\Delta \bar u(\tilde\x)=&\, -\frac{\lambda \exp \left\{-\ds\bar  u(\tilde\x) \right\}}{\ds\int_{\tilde\Omega} \exp \left\{-\ds
	\bar u( \s)\right\}\, d\s}\label{eqsymm}
\end{align}
and the boundary condition \eqref{compatibility} becomes
\beq\label{Boundary_0}
\frac{\p \bar u(\tilde\x)}{\p n}=-\frac{\lambda}{|\p \tilde\Omega|}\hspace{0.5em} \mbox{for}\ \tilde\x\in\p
\tilde\Omega.
\eeq
We consider now the PNP problem \eqref{eqsymm}-\eqref{Boundary_0} in the solid of revolution (Fig. \ref{f:conf}A), obtained by rotating the symmetric planar domain Fig. \ref{f:conf}B around its $z-$axis of symmetry. Consequently, $\tilde\Omega$ represents now a ball with a cusp-shaped funnel, with a radius curvature $R_c$ at the entrance of the funnel (blue dashed circles in Fig. \ref{f:conf}A-B).

Using the change of variable $\ds \x=\frac{\tilde\x}{R_c}$, $\ds\p\Omega=\frac{\p\tilde\Omega}{{R_c}^{2}}$ and $\ds \Omega=\frac{ \tilde\Omega}{{R_c}^{3}}$ and $u(\x)=\bar u(\x)+{\ds\ln\left(\lambda R_c^2/\int_{\tilde\Omega}\exp\{-u(\s)\}\,d\s \right)}$ converts \eqref{eqsymm}  into
\begin{align}\label{NewPb}
-\Delta {u}(\x)=&\,\exp\{-{u}(\x)\}\hspace{0.5em}\mbox{for}\ \x\in \Omega\\
\frac{\p {u}(\x)}{\p n}=&\,-\frac{\lambda }{|\p  \Omega|R_c}\hspace{0.5em}\mbox{for} \ \x\in\p \Omega.\nonumber
\end{align}
The non-dimensional surface charge density is
\beq
\sigma &=& \frac{\lambda }{|\p  \Omega|R_c }.
\eeq
We first consider a uniform surface charge density in \eqref{NewPb} and then study the consequences of a non-homogeneously distributions.
\subsection{Poisson-Nernst-Planck solutions in a 3D cusp-shaped funnel}\label{s:Reduction_cplx}
The cylindrical symmetry of the Neumann boundary value problem (BVP) \eqref{NewPb} in the $(r,z,\phi)$  coordinates (Fig. \ref{f:conf}A) centered on the axis of symmetry, implies that $\tilde u(\x)$ is independent of the angle $\phi$ in the domain $\Omega$. It follows that \eqref{NewPb} in the domain $\Omega$ can be written as
\beq\label{Problem_Cusp}
\frac{\p^2 u(r,z)}{\p r^2}+\frac{1}{r}\frac{\p u(r,z)}{\p r }+\frac{\p^2 u(r,z)}{\p z^2}&=&-\exp(-u(r,z))\\
\frac{\p u(r,z)}{\p n}&=&-\sigma,\nonumber
\eeq
where $n=[n_r,n_z]^T$ is the outward normal unit vector to the surface $\p \Omega$ and $r$ is the distance to the symmetry axis of $\Omega$. 
The opening at the cusp funnel is small $\overline{AB}= \eps\ll1 $ (green line Fig. \ref{f:conf}B), so the funnel is a narrow passage.
To remove the cusp singularity, we use first the transformation to the rotated and translated coordinates given by $\ds\tilde{r}=r-1-\eps/2$ and $\tilde z = -z+1$. Setting $u(r,z)=\tilde u ( \tilde r, \tilde z )$, eq. \eqref{Problem_Cusp} becomes,
\beq\label{Problem_Cusp2}
\frac{\p^2 \tilde  u(\tilde r,\tilde z)}{\p \tilde r^2}+\frac{\p^2  \tilde u(\tilde r,\tilde z)}{\p \tilde z^2}+\frac{1}{(\tilde r+1+\eps/2)}\,\frac{\p  \tilde u(\tilde r,\tilde z)}{\p \tilde r }&=&-\exp(- \tilde u(\tilde r,\tilde z))\\
\frac{\p \tilde  u(\tilde r,\tilde z)}{\p \tilde n}&=&-\sigma.\nonumber
\eeq
We shall construct an asymptotic expansion of the solution $ \tilde u(\tilde r,\tilde z)$ for small $\eps$ by first mapping the cross section in the $(\tilde r,\tilde z)-$plane conformally into its image under the M\"obius transformation \cite{HS2012}
\beq\label{Mobius}
w(\xi)&=&\rho e^{i\theta}=\frac{\xi-\alpha}{1-\alpha\xi},
\eeq
where
\beq
\alpha &=& -1-\sqrt{\eps}+O(\eps),
\eeq
and $\xi = \tilde r +i \tilde z$.  In the dimensionless domain $\Omega$, the parameter $\eps$ is also dimensionless and $R_c\eps=\tilde \eps$. M\"obius transformation maps the two osculating circles $A$ and $B$ (dashed blue) into concentric circles (see
Fig. \ref{f:conf}B-C). The M\"obius transformation \eqref{Mobius} maps the right circle $B$ (dashed blue) into itself and $\Omega$ is mapped onto the banana-shaped domain $\Omega_w=w(\Omega)$ as shown in Figure \ref{f:conf}C.\\
The second order derivative for $\tilde{u}(\xi) = v(w)$ is computed using \eqref{Mobius} in \eqref{Problem_Cusp2} \cite{Henricci}
\beq\label{Trans_delta}
\frac{\p^2 \tilde  u}{\p \tilde r^2}+\frac{\p^2  \tilde u}{\p \tilde z^2}&=& |w'(\xi)|^2 \Delta_w  v(w).
\eeq
In the small $\eps$ limit, we have
\beq
|w'(\xi)|^2&=&\frac{|(1-\sqrt{\eps})e^{i\theta}-1+O(\eps)|^4}{4\eps +O(\eps^{3/2})}.
\eeq
The 3D BVP \eqref{Problem_Cusp2} differs from the 2D problem \cite{NonLin2017} by the extra first order radial derivative. For small $\eps$ limit, we have
\beq\label{tilde_r}
\tilde r +1+ \eps/2&=& \frac{\eps}{1-\cos(\theta)} +O(\eps^{3/2}) .
\eeq
In complex coordinates we have
\beq\label{diff_r}
\frac{\p u( \tilde r, \tilde z)}{\p \tilde r} &=& \Re e \left(  \nabla u( \xi)\right),
\eeq
where $\Re e(\cdot)$ is the real part. Under the conformal mapping \eqref{Mobius}, the gradient from \eqref{diff_r} transforms as follows \cite{Henricci}
\beq\label{grad_trans}
\nabla u(\xi) &=& \nabla_w v(w) \, \overline{w'(\xi)}.
\eeq
Using polar coordinates $(\rho,\theta)$ in the mapped domain $\Omega_w$, we write
\beq\label{w1_w2}
\overline{w'(\xi)}&=&w_1(\rho,\theta)+i\,w_2(\rho,\theta),
\eeq
where $i^2=-1$.
Using \eqref{Mobius}, we obtain
\beq\label{w1_w2_2}
\tilde w_1(\rho,\theta)&=& \frac{1-\alpha^2 \rho^2 +2\alpha \rho \cos(\theta)(1+ \alpha  \rho   \cos(\theta)) }{1-\alpha^2}\\
\nonumber  \\
\tilde w_2(\rho,\theta)&=& -2\alpha \rho\sin(\theta)\frac{ 1 + \alpha  \rho   \cos(\theta) }{1-\alpha^2}. \nonumber
\eeq
Using \eqref{diff_r} and \eqref{w1_w2_2}, in polar coordinates (see Appendix), it follows that
\beq\label{diff_r3}
\frac{\p \tilde u(\tilde r, \tilde z)}{\p \tilde r} &=& \frac{\p  \tilde v(\rho,\theta)}{\p \rho }\left( \cos(\theta) \tilde w_1(\rho,\theta)  - \sin(\theta)\tilde w_2(\rho,\theta)  \right) \\
&&-\frac{1}{\rho}\frac{\p  \tilde v(\rho,\theta) }{\p \theta }\left( \sin(\theta)\tilde w_1(\rho,\theta) +\cos(\theta) \tilde w_2(\rho,\theta) \right). \nonumber
\eeq
To leading order, using \eqref{tilde_r} and \eqref{diff_r3}, we get (Appendix)
\beq\label{Diff_order_1}
\frac{1}{\tilde r}\frac{\p \tilde u(\tilde r,\tilde z)}{\p \tilde r}&=&
 -  \frac{ \rho(1-\cos(\theta))^2}{\eps^{3/2}}\frac{\p\tilde v (\rho,\theta )}{\p\rho}
- \frac{\sin(\theta)(1-\cos(\theta))}{\eps}\frac{\p\tilde v (\rho,\theta  )}{\p\theta}. 
\eeq
In summary, using  \eqref{Trans_delta} in polar $(\rho,\theta)-$coordinates,  eq. \eqref{Diff_order_1} and \eqref{Problem_Cusp2} in $\Omega_w$, are changed to
{\small
\beq\label{Problem_banana_1}
\frac{|(1-\sqrt{\eps})e^{i\theta}-1|^4}{4\eps}\left(\frac{\p^2 \tilde v(\rho,\theta)}{\p \rho^2}+\frac{1}{\rho}\frac{\p\tilde v(\rho,\theta)}{\p \rho}+\frac{1}{\rho^2}\frac{\p^2\tilde v(\rho,\theta)}{\p \theta^2} \right) &-&\frac{ \rho(1-\cos(\theta))^2}{\eps^{3/2}}\frac{\p\tilde v (\rho,\theta  )}{\p\rho} \nonumber\\
&-&\frac{\sin(\theta)(1-\cos(\theta))}{\eps}\frac{\p\tilde v (\rho,\theta  )}{\p\theta}\nonumber\\
&=&-\exp\left\{-\tilde v(\rho,\theta) \right\}\nonumber\\
\frac{\p\tilde v(\rho,\theta) }{\p n}&=&-\frac{\sigma\sqrt{\eps}}{  1-\cos(\theta) }.
\eeq
}
\subsection{Asymptotic analysis of the PNP equations in a cusp-shaped funnel}\label{s:Cusp_3D}
\begin{figure}[http!]
	\center
	\includegraphics[scale=0.11]{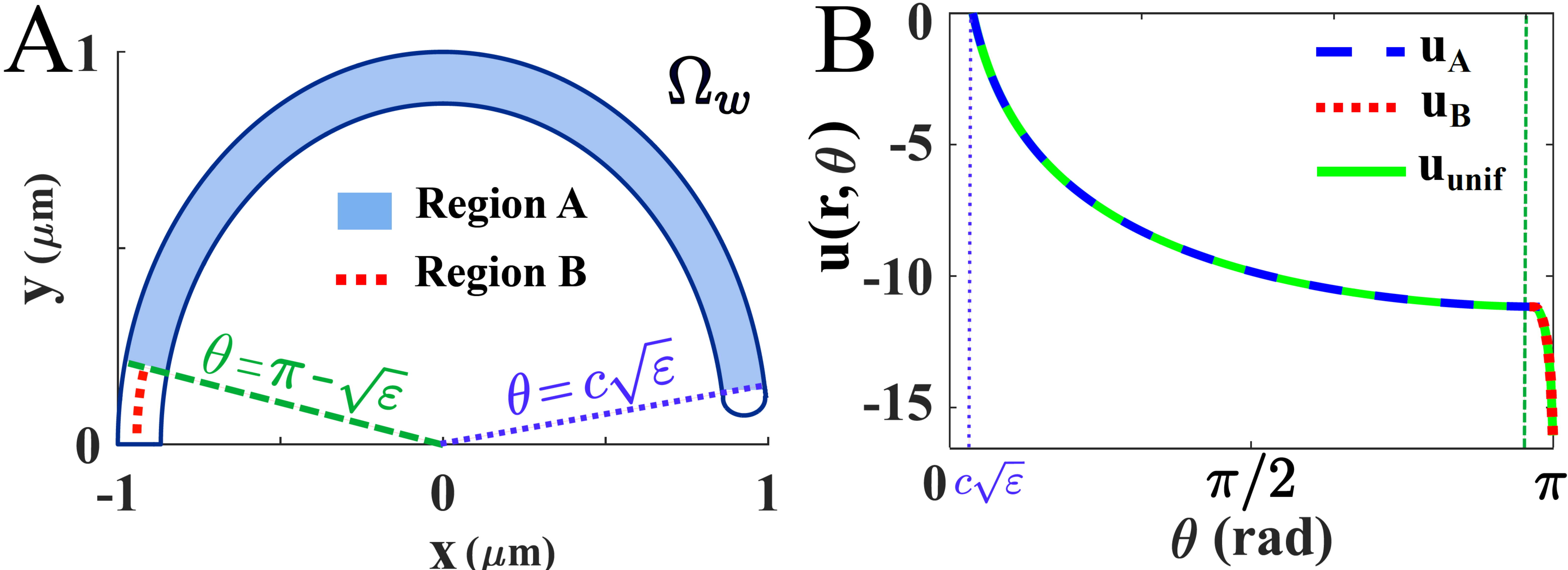}
	\caption{\small  { {\bf Decomposition of the domain $\Omega_w$ into two subregions regions $A$ and $B$ }
			{\bf A.} Representation of the two subregions
			$A$ (blue) and $B$ (dotted red) of $\Omega_w$.
			{\bf B.} Solutions of \eqref{Solution_uA_theta} (dashed blue), \eqref{uB} (red dots), and the uniform approximation  $u_{unif}$ \eqref{Sol_unif} (green) for $r=1-\sqrt{\eps}$.}}
	\label{f:RegionsAB}
\end{figure}

To analyse eq. \eqref{Problem_banana_1} in the limit of $\sigma \gg1$, $\eps\rightarrow 0$ \cite{PhysD2016}, we approximate the domain $\Omega_w$ by two subregions
\beq
A&=& \{ (\rho,\theta)\in \Omega_w \,: \, |\theta-\sqrt{\eps}|>\pi,\, |\rho-1|\leq\sqrt{\eps}  \}\\
\nonumber\\
B&=& \{ w=(1-\sqrt{\eps})e^{i\theta}\,: \, |\theta-\pi|\leq\sqrt{\eps}\}, \nonumber
\eeq
as  illustrated in Fig. \ref{f:RegionsAB}A. The regions $B$ consists of a circular arc (dashed red).  We construct now the solution $u_A(r, \theta)$ and $u_B(\theta)$ of \eqref{NewPb} in each subregion.
\subsubsection*{Asymptotics of $u_A(r,\theta)$ in region $A$}\label{s:uA}
To construct the asymptotics solution $u_A(r,\theta)$ in region $A$, we use that the radial derivative $\frac{\p}{\p r}$ is $O(\sigma\sqrt{\eps})\to\infty$ in the regime $\sigma\eps^{3/2}=O(1)$ as $\sigma\to\infty$ and $\eps\to0$. Thus the angular derivatives are negligible relative to the radial ones.  It follows in a regular expansion of the solution, the $\theta$ derivative can be neglected relative to the $\rho$ derivative and we will equation \ref{Problem_banana_1} along the rays $\theta=\theta_0=const$, for $\rho\in[1-\sqrt{\eps},1]$.

Setting $u_A(\rho,\theta_0)=v(\rho, \theta_0)$, to leading order in $\sigma\sqrt{\eps}$, equation \eqref{Problem_banana_1} reduces to
\beq\label{Pb_uA}
-e^ {-\ds u_A(\rho,\theta_0)  }&=&\frac{|(1-\sqrt{\eps})e^{i\theta_0}-1|^4}{4\eps}\left(\frac{\p^2 u_A(\rho,\theta_0)}{\p \rho^2}+\frac{1}{\rho}\frac{\p u_A(\rho,\theta_0)}{\p \rho}  \right)\\
&&  -\frac{ \rho(1-\cos(\theta_0))^2}{\eps^{3/2}}\frac{\p\tilde u_A(\rho,\theta_0)}{\p\rho}\nonumber\\
\left.\frac{d u_A(\rho,\theta_0)}{d \rho}\right|_{\rho=1} &=& -\frac{ \sqrt{\eps} }{ 1-\cos(\theta_0)}\nonumber \\
\left.\frac{d u_A(\rho,\theta_0)}{d \rho}\right|_{\rho=1-\sqrt{\eps}} &=& 0.\nonumber
\eeq
In the limit $\eps\ll1$, we note that $ |\rho e^{i\theta_0}(1-\sqrt{\eps})-1|^4=|e^{i\theta_0}-1|^4+O(\sqrt{\eps})$ and
using the change of variable $\rho = \tilde \rho \sqrt{\eps}$ and setting $ u_A(\rho,\theta_0)= v_A(\tilde \rho,\theta_0)$, to leading order in $\eps\ll1$, eq. \eqref{Pb_uA} becomes
\beq\label{Reduced_uA0}
-\frac{4\eps^2e^ {-\ds  v_A(\tilde \rho,\theta_0) }}{|e^{i\theta_0}-1|^4}&=& \frac{\p^2   v_A(\tilde \rho,\theta_0)}{\p  \tilde\rho^2}-\sqrt{\eps}\frac{\p   v_A(\tilde \rho,\theta_0)}{\p \tilde\rho}\left( 1-\frac{ 4(1-\cos(\theta_0))^2 }{|e^{i\theta_0}-1|^4}\right).
\eeq
Using the function,
\beq\label{h0}
h(\theta_0)&=&\frac{4\eps^2 }{|e^{i\theta_0}-1|^4}
\eeq
and $\tilde v_A(\tilde \rho,\theta_0)=v_A(\tilde \rho,\theta_0)-\ln(h(\theta_0))$, eq. \eqref{Reduced_uA0} is transformed into
\beq\label{Reduced_uA2}
\frac{\p^2  \tilde v_A(\tilde \rho,\theta_0)}{\p  \tilde\rho^2}&=&-e^ {-\ds  v_A(\tilde \rho,\theta_0) } +\sqrt{\eps}\frac{\p   v_A(\tilde \rho,\theta_0)}{\p \tilde\rho}\left( 1-\frac{ (1-\cos(\theta_0))^2 }{|e^{i\theta_0}-1|^4}\right).
\eeq
\begin{figure}[H]
	\center
	\includegraphics[scale=0.73]{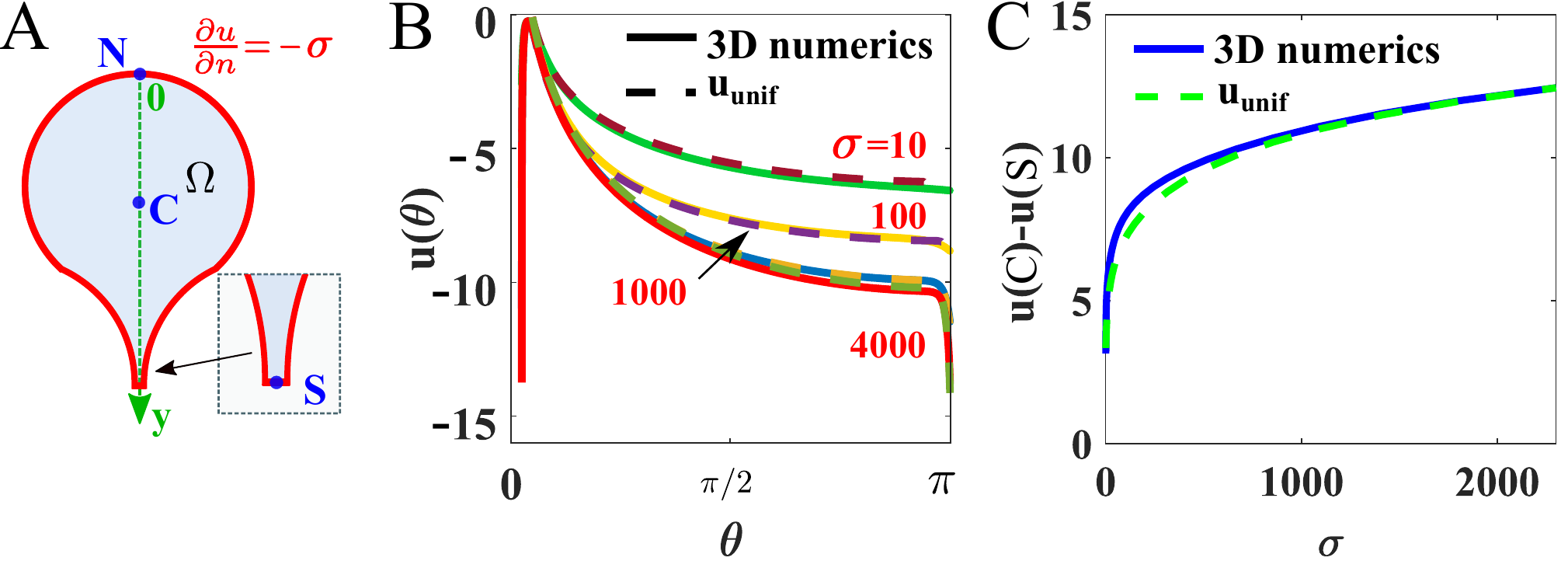}
	\caption{ {\small  {\bf PNP solution \eqref{NewPb} in a 3D domain with a cusp-shaped funnel}
		{\bf A.} Representation of the domain $\Omega$ with a surface charge density $\sigma$, the north pole $N$, the funnel tip $S$, and the center of mass $C$, respectively.
		{\bf B. } Numerical \eqref{NewPb} (solid) and analytical \eqref{Sol_unif} (dashed) solutions in the domain $\Omega_w$ for several values of $\sigma=10,\, 100,\,  1000$ and $4000$ for $\eps=0.01$.
		{\bf C.} Difference $u(C)-u(S)$ computed numerically (solid blue) from \eqref{NewPb} and analytically (dashed green) from \eqref{unif_diff}.   }\label{f:3D_Cusp}}
\end{figure}
Using a regular expansion in the small $\eps$ limit (in the regime $\sigma\eps^{3/2}=O(1)$)
\beq\label{Reg_expansion}
\tilde v_A(\tilde \rho,\theta_0)&=&\tilde v_{A,0}(\tilde \rho,\theta_0)+\sqrt{\eps}\tilde v_{A,1}(\tilde \rho,\theta_0)+O(\eps)
\eeq
in \eqref{Reduced_uA2}, we get
\beq\label{Leading_uA}
\frac{\p^2\tilde  v_{A,0}(\tilde \rho,\theta_0)}{\p  \tilde\rho^2}&=&-e^ {-\ds \tilde v_{A,0}(\tilde \rho,\theta_0) }  \\
\left.\frac{\p   \tilde v_{A,0}(\tilde \rho,\theta_0)}{\p\tilde \rho}\right|_{\tilde\rho=0} &=&  \frac{\sigma \eps  }{  1-\cos(\theta_0) }\nonumber \\
\left.\frac{\p    \tilde v_{A,0}(\tilde \rho,\theta_0)}{\p \tilde\rho}\right|_{\tilde \rho=1 } &=& 0.\nonumber
\eeq
A direct integration of \eqref{Leading_uA} is \cite{NonLin2017}
\beq\label{General_Sol}
\tilde v_{A,0}(\tilde \rho,\theta_0)&=& \ln\left( 2C_1(\theta_0)^2\cos^2\left(    \frac{\tilde \rho + C_2(\theta_0)}{2C_1(\theta_0)}\right)\right),
\eeq
where $C_1(\theta_0)$ and $C_2(\theta_0)$ are two constants that depend on $\theta_0$.  To compute these constants, we differentiate \eqref{General_Sol}
\beq\label{diff_sol_A}
\tilde v'_{A,0}(\tilde \rho,\theta_0)&=&\frac{-1}{C_1(\theta_0)}\tan\left(\frac{\tilde  \rho+C_2(\theta_0) }{2C_1(\theta_0)}\right).
\eeq
Using the Neumann boundary condition at $\tilde \rho=1$ in \eqref{Leading_uA}, we get
\beq\label{val_C2}
C_2(\theta_0)&=&-1.
\eeq
Using \eqref{val_C2} and \eqref{diff_sol_A} and the boundary condition at $\tilde \rho=0$ in \eqref{Leading_uA}, we find that $C_1$ is solution of the transcendental equation,
\beq
\frac{\sigma\eps C_1(\theta_0) }{ (1-\cos(\theta_0))}&=&\tan\left(\frac{1}{2C_1(\theta_0)}\right).
\eeq
In the regime $\sigma=O(\eps^{-3/2})$, we have
\beq\label{C1_const}
C_1(\theta_0)&=&\frac{2 (1-\cos(\theta_0)) +\sigma\eps}{\pi\sigma\eps} + O\left(\frac{1}{\sigma\eps}\right).
\eeq
Using \eqref{General_Sol},\eqref{val_C2} and \eqref{C1_const} in \eqref{General_Sol}, we obtain to leading order
\beq\label{Sol_VA0}
\tilde v_{A,0}(\tilde \rho,\theta_0)&=& \ln\left( 2\left(\frac{2 (1-\cos(\theta_0)) +\sigma\eps}{\pi\sigma\eps}\right)^2\right)\\
&&+\ln\left(\cos^2\left(    \frac{\pi\sigma\eps(\tilde \rho -1)}{2(2 (1-\cos(\theta_0))+\sigma\eps)}\right)\right).\nonumber
\eeq
Using \eqref{Sol_VA0}, \eqref{Reg_expansion} and \eqref{h0}, we conclude
\beq\label{Solution_In_A}
v_A(\tilde \rho,\theta_0)&=& \ln\left( 2\left(\frac{2 (1-\cos(\theta_0)) +\sigma\eps}{\pi\sigma\eps}\right)^2\right)+\ln\left(\frac{4\eps^2 }{|e^{i\theta_0}-1|^4}\right)\\
&&+\ln\left(\cos^2\left(    \frac{\pi\sigma\eps(\tilde \rho -1)}{2(2 |(1-\cos(\theta_0))+\sigma\eps)}\right)\right)\nonumber \\
&& +O(\sqrt{\eps}).\nonumber
\eeq
In particular the solution at $\rho=1-\sqrt{\eps}$ is
\beq\label{Solution_uA_theta}
u_A( 1-\sqrt{\eps},\theta_0)&=& \ln\left( 8\left(\frac{2 (1-\cos(\theta_0)) +\sigma\eps}{\pi\sigma|e^{i\theta_0}-1|^2}\right)^2\right)+O(\sqrt{\eps}).
\eeq
We note that the three dimensional solution \eqref{Solution_uA_theta} is identical to the one obtained inside a planar cusped-shaped domain \cite{NonLin2017}.
\subsubsection*{Asymptotics of $u_B(\theta)$ in region $B$}\label{s:uB}
The asymptotic solution $u_A(\rho,\theta)$ in $A$ does not satisfy the boundary condition \eqref{Problem_banana_1} at $\theta=\pi$. Indeed, $\p u_A(\rho,\theta)/\p\theta|_{\theta=\pi} =0$, while the boundary condition  \eqref{Problem_banana_1} is $\p v/\p \theta|_{\theta=\pi}= -\sigma\sqrt{\eps}/2 \gg1$, thus a boundary layer should develop.

The boundary layer solution $u_B(\theta)$ is derived  by taking into account the $\theta$ derivatives in eq. \eqref{Problem_banana_1}:
\beq\label{Problem_banana_B}
\frac{|(1-\sqrt{\eps})e^{i\theta}-1|^4}{4\rho^2\eps} \frac{\p^2 u_B(\theta)}{\p \theta^2}
+ \frac{\sin(\theta)(1-\cos(\theta))}{\eps}\frac{\p\tilde u_B(\theta)}{\p\theta}&=&-e^{-\ds u_B(\theta) }.
\eeq
In small $\eps$ limit, for $\rho=1-\sqrt{\eps}$, we have
\beq\label{approx_B}
\frac{ 4\eps}{|\rho e^{i\theta}(1-\sqrt{\eps})-1|^4}&=&\frac{ \eps}{4},
\eeq
which is constant. Using \eqref{approx_B} in \eqref{Problem_banana_B} and $\eta=\pi-\theta$, we define $u_B(\theta)=\tilde u_B(\eta)$, leading to
\beq\label{Problem_banana_B2}
 \frac{\p^2 \tilde u_B(\eta)}{\p \eta^2}
- \frac{1}{4}\sin(\eta)(1+\cos(\eta))\frac{\p\tilde u_B(\theta)}{\p\eta}&=&-\frac{\eps}{4}e^{-\ds \tilde u_B(\eta) }.
\eeq
Since $0\le\eta\le\sqrt{\eps}$, we shall approximate the first order term and thus eq. \eqref{Problem_banana_B2} reduces to
\beq\label{Problem_banana_B3}
\frac{\p^2 \tilde u_B(\eta)}{\p \eta^2}
- \frac{\eta}{2}\frac{\p\tilde u_B(\theta)}{\p\eta}&=& -\frac{\eps}{4}e^{-\ds \tilde u_B(\eta) }.
\eeq
Using $v(\eta)=u_B(\eta)-\ln{\left(\ds4/\eps\right)}$, eq. \eqref{Problem_banana_B3} is transformed to
\beq\label{Problem_banana_B4}
 -\frac{\p^2 \tilde v(\eta)}{\p \eta^2}
+ \frac{\eta}{2}\frac{\p\tilde v(\eta)}{\p\eta}&=&e^{-\ds \tilde v(\eta) }.
\eeq
Using the boundary condition \eqref{Problem_banana_1}, we further reduce the solution $v(\eta)$ to the equation
\beq\label{Problem_banana_B5}
 -\frac{\p^2 \tilde v(\eta)}{\p \eta^2}&=& e^{-\ds \tilde v(\eta) }
+ O(\lambda\eps^2)\\
\left.\frac{\p v(\eta)}{\p\eta}\right|_{\eta=0}&=& \frac{ \sigma\sqrt{\eps}}{2 }\nonumber\\
\left.\frac{\p v(\eta)}{\p\eta}\right|_{\eta=\sqrt{\eps}}&=& 0.\nonumber
\eeq
The solution is
\beq\label{General_Sol2}
  \tilde v(\eta) &=& \ln\left( 2\tilde C_1 ^2\cos^2\left(    \frac{  \eta + \tilde C_2 }{2\tilde C_1 }\right)\right),
\eeq
where
\beq\label{C2_B}
\tilde C_2&=&-\sqrt{\eps},
\eeq
and $\tilde C_1$ is solution of the transcendental equation
\beq
\frac{2\tilde C_1}{\sqrt{\eps}}\arctan\left( \frac{\sigma\sqrt{\eps}\tilde C_1 }{2 }\right)&=&1.
\eeq
In the limit $\sigma\gg 1$, we have
\beq\label{C1_B}
\tilde C_1&=&\frac{2}{\pi }\left( \frac{\sqrt{\eps}}{2} + \frac{2  }{\sigma \sqrt{\eps}}  \right)+O\left(\frac{1}{(\sigma \sqrt{\eps})^3} \right).
\eeq
We note that $\ds\frac{\eta}{2}\frac{\p\tilde v(\eta)}{\p\eta}$ is small, justifying our simplifications.
We conclude from \eqref{C1_B}-\eqref{C2_B}-\eqref{General_Sol} that for $\theta \in B$, the asymptotic solution is
\beq\label{uB}
u_{B}(\theta)&=& \ln\cos^2 \frac{\pi}{2}\sqrt{\frac{(\theta-(\pi-\sqrt{\eps}))^2}{\eps}}
\left(1-\frac{4 }{\sigma\eps} \right) +C_0,
\eeq
where $C_0$ is a constant that we find in the next paragraph by matching the solution in two regions $A$ and $B$.
\subsubsection*{A uniform approximation of $u(\rho,\theta)$ in $\Omega_w$}\label{s:unif}
We now construct a uniform asymptotic approximation $u_{unif}(\rho,\theta)$ in the region $A\cup B$ (Fig. \ref{f:3D_Cusp}A) using  $u_A(\rho,\theta)$ with $u_B(\rho,\theta)$ that match for $\theta=\pi-\sqrt{\eps}$, leading to
\beq
C_0&=&u_A\left(1-\sqrt{\eps},\pi-\sqrt{\eps}\right).
\eeq
Using the analytical expression \eqref{Solution_uA_theta} of $u_A$, we get
\beq\label{C_0}
C_0&=& \ln\left(  \frac{\left(4  +\sigma\eps\right)^2}{2(\pi\sigma )^2}\right).
\eeq
Thus,
\beq\label{uB_C0}
u_{B}(\theta)&=& \ln\cos^2 \frac{\pi}{2}\sqrt{\frac{(\theta-(\pi-\sqrt{\eps}))^2}{\eps}}
\left(1-\frac{4 }{\sigma\eps} \right) + \ln\left(  \frac{\left(4   +\sigma\eps\right)^2}{2(\pi\sigma )^2}\right).
\eeq
Consequently, using \eqref{Solution_uA_theta} and \eqref{uB_C0} the solution in the funnel is
\beq\label{Sol_unif}
u_{unif}(\rho,\theta)&=&
{\small
\begin{cases}
	   \ds \ln\left( 8\left(\frac{2 (1-\cos(\theta)) +\sigma\eps}{\pi\sigma|e^{i\theta}-1|^2}\right)^2\right), &\hbox{ for }  \theta \in [0, \pi-\sqrt{\eps}] \\
  \\
	 \ds\ln\cos^2 \frac{\pi}{2}\sqrt{\frac{(\theta-(\pi-\sqrt{\eps}))^2}{\eps}}
	\left(1-\frac{4 }{\sigma\eps} \right) + \ln\left(  \frac{\left(4   +\sigma\eps\right)^2}{2(\pi\sigma )^2}\right),& \hbox{ for }  \theta \in [\pi-\sqrt{\eps},\pi].
\end{cases}}
\eeq
The numerical solution of eq. \eqref{NewPb} in $\Omega_w$ and the approximation $u_{unif}(\rho,\theta)$ of \eqref{Sol_unif} are shown in Fig. \ref{f:3D_Cusp}B.
\subsection{Estimating the potential drop in $\Omega_w$}
The difference of potential between the center of mass $C$ and the tip of the funnel $S$ (see Fig. \ref{f:3D_Cusp}A) is defined as
\beq
\tilde\Delta_{funnel} u=u(C)-u(S),
\eeq
where
\beq
u(S)=u(1-\sqrt{\eps},\pi)\quad\mbox{and}\quad u(C)=u(1-\sqrt{\eps},c\sqrt{\eps}),
\eeq
$u$ is solution of eq. \ref{NewPb} and the constant $c$ depends on the domain geometry  and is defined by the conformal mapping $w$ (relation \eqref{Mobius}). To compute $\tilde \Delta_{funnel} u$, we use the two differences
\beq\label{diff_A_exp}
\tilde\Delta u_A &=&u_A(1-\sqrt{\eps},\pi) -u_A(1-\sqrt{\eps},c\sqrt{\eps}),
\eeq
and
\beq\label{diff_B_exp}
\tilde \Delta u_{B}=u_B (\pi)-u_B (\pi-\sqrt{\eps}).
\eeq
It follows that
\beq\label{D_unif}
\tilde \Delta_{funnel} &=& \tilde \Delta u_{A}+\tilde \Delta u_{B}.
\eeq
To compute $\tilde \Delta u_{A}$, we use the analytical expression \eqref{Solution_uA_theta} for $\rho=1-\sqrt{\eps}$ and any $\theta_0$,
\beq\label{v_regA1}
u_{A}(1-\sqrt{\eps},\theta_0)=-\ln\frac{|e^{i\theta_0}-1|^4}{8(1-\sqrt{\eps})^2}
\left(\frac{\sigma\pi}{2 (1-\cos(\theta_0))+\sigma\eps}\right)^2+O(\eps).
\eeq
At the point $S$ ($\theta_0=\pi$),
\beq\label{solution4pi_v1}
u_A(S)=-\ln\frac{2\sigma^2\pi^2}{(4 +\sigma\eps)^2} +2\ln(1-\sqrt{\eps})+O(\eps).
\eeq
To estimate $u_A(C)$ for which $\theta_0=c\sqrt{\eps}$, we observe that for $\eps\ll1$ in relation \eqref{v_regA1},
\beq\label{expan01}
|e^{i\theta_0}-1|^4=c^4\eps^2+O(\eps^3),
\eeq
and
\beq\label{expan02}
2(1-\cos (c\sqrt{\eps}))+\sigma\eps&=&\eps( c^2+\sigma)+O(\eps^2).
\eeq
We use \eqref{expan01} and \eqref{expan02}, so eq. \eqref{v_regA1} reduces to
\beq\label{solution4c_v0}
u_A(C)={-\ln \frac{c^4} {8}  \left(\frac{\sigma\pi}{ c^2+\sigma} \right)^2}+2\ln(1-\sqrt{\eps})+O\left(\eps\right).
\eeq
In the large $\sigma$ limit,
\beq\label{solution4c_v1}
u_A(C)={-\ln \frac{\pi^2c^4}{8}}+2\ln(1-\sqrt{\eps})+O\left(\eps,\frac{1}{\sigma}\right).
\eeq
Using $u_A(C)$ and $u_A(S)$, we conclude that
\beq\label{diff_A}
\tilde\Delta u_A =-\ln\frac{2\sigma^2\pi^2}{(4 +\sigma\eps)^2}+\ln\frac{\pi^2c^4}{8} +O\left(\eps,\frac{1}{\sigma}\right).
\eeq
For $\sigma\gg1$, to leading order, the solution of eq. \eqref{diff_A} does not depend on $\sigma$
\beq\label{Diff_A_large}
\tilde \Delta u_A\sim{-\ln\frac{2^4}{c^4\eps^2}}.
\eeq
We now estimate the difference $ \tilde\Delta u_{B}$. We have from \eqref{uB} that
\beq\label{ub_eps}
u_B(\pi-\sqrt{\eps})=C_0
\eeq
and
\beq\label{ub_pi}
u_B(\pi)=  \ln \sin^2\left(\frac{\pi}{\sigma\eps} \right) +C_0.
\eeq
Using \eqref{ub_eps} and \eqref{ub_pi} in \eqref{diff_B_exp}, {we obtain
\beq\label{diff_bl}
 \tilde\Delta u_{B}= \ln \sin^2 \left(\frac{\pi}{\sigma\eps} \right).
 \eeq
For $\sigma\gg1$, eq. \eqref{diff_bl} reduces to
 \beq\label{diff_B_large}
 \tilde\Delta u_{B}=-2\ln\sigma+2\ln\frac{ \pi}{\eps}+O\left(\frac{1}{\sigma^2}\right).
 \eeq
 Finally, using \eqref{diff_A}, \eqref{diff_bl} and \eqref{D_unif}, we find that the difference in the funnel is
 \begin{align}\label{unif_diff}
 \tilde\Delta u=&\ln\sin^2\frac{\pi }{\sigma\eps}-\ln\frac{2
 	\sigma^2\pi^2}{(4 +\sigma\eps)^2}+\ln\frac{\pi^2c^4}{8} \,+ O\left(\eps,\frac{1}{\sigma}\right).
 \end{align}
The results in large $\sigma$ limit found in \eqref{Diff_A_large}, \eqref{diff_B_large} and leads to
 \beq\label{diff_unif2}
 \tilde\Delta u=-\ln\sigma^2+2\ln\frac{\pi c^2 }{4}+O\left(\frac{1}{\sigma}\right).
 \eeq
 Equation \eqref{diff_bl} shows that for $\sigma\gg1$, the potential drop in the cusp-shaped funnel is dominant in region $B$. We compare (Fig. \ref{f:3D_Cusp}C) expression \eqref{unif_diff} with the numerical solution of \ref{NewPb}. We note that the distribution of the potential inside a 3D solid funnel is to leading order identical to the one we obtained inside a planar cusp \cite{NonLin2017}.

\section{The PNP equations in a cusp-shaped domain with non-homogeneous surface charge density}\label{s:non_uniform}
When the surface charge density is not homogeneously distributed over the surface $\p\Omega$, we expect a re-organization of the potential $u$ of  \eqref{NewPb}.
we subdivide the surface $\p\Omega$ into three regions (Fig. \ref{f:split_bdv}),
\beq\label{split_surf}
\p\Omega &=&\p\Omega_{\eps}\cup\p\Omega_{cusp}\cup\p\Omega_{bulk},
\eeq
where $\p\Omega_{\eps}$ is the bottom of the funnel, $\p\Omega_{cusp}$ the funnel area and $\p\Omega_{bulk}$ the bulk surface. The Neuman boundary conditions on each sub-regions are defined by
\beq\label{2_BL_conditions}
\frac{\p u(\x)}{\p n}&=& \frac{-\lambda_{\eps} }{|\p\Omega_{\eps}|}\mbox{ on } \p\Omega_{\eps}\\
\frac{\p u(\x)}{\p n}&=& -\frac{ \lambda_{cusp} }{|\p\Omega_{cusp}|}\mbox{ on } \p\Omega_{cusp} \nonumber
\\
\frac{\p u(\x)}{\p n}&=& \frac{ -\lambda_{bulk} }{|\p\Omega_{bulk}|}\mbox{ on } \p\Omega_{bulk}.\nonumber
\eeq
Using the compatibility condition obtained by integrating the Poisson equation \eqref{NewPb}
\beq\label{Compatibility}
\int_{\p\Omega}\frac{\p u(\x) }{\p n}  dS&=& -\lambda.
\eeq
we obtain that
\beq\label{lambda_split}
\lambda&=&\lambda_{\eps}+\lambda_{cusp}+\lambda_{bulk}.
\eeq
We will use the notation
\beq\label{sigma_notation}
\sigma_j=\ds\frac{ \lambda_{j} }{|\p\Omega_{j}|},
\eeq
where $j\in \{ \eps\,,\, cusp\,,\, bulk\}$.
\begin{figure}[H]
	\center
	\includegraphics[scale=0.75]{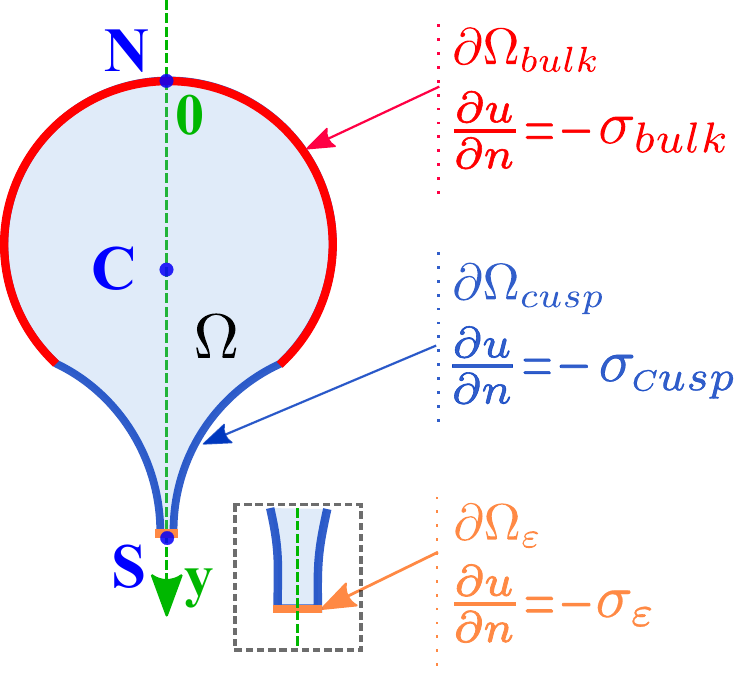}
	\caption{ {\small  {\bf Schematic representation of the $\p\Omega$ boundary subregions.}
 Subregions of the boundary $\p\Omega$: the cusp $\p\Omega_{cusp}$ (red), the bulk $\p\Omega_{bulk}$ (blue) and (as shown in the inset panel) the funnel bottom $\p\Omega_{\eps}$ (orange). Their respective surface charge densities are $\sigma_{bulk}$, $\sigma_{cusp}$ and $\sigma_{\eps}$.
		}\label{f:split_bdv}}
\end{figure}
\subsection{PNP solutions for $\sigma_{cusp}=0,\sigma_{bulk}=\sigma_{\eps}=\sigma$ in 3D }\label{s:Null}
To compute the solution of \eqref{NewPb} for an uncharged funnel ($\sigma_{cusp}=0$), we will use the same conformal mapping \eqref{Mobius} as describe above with now reflecting boundary condition on $\p\Omega_{cusp}$, which are invariant under the conformal mapping. As a result the boundary conditions on the two like-style arcs of the domain $\Omega_w$ are also reflective. Consequently, instead of searching a solution in the banana-shaped domain $\Omega_{w}$, we will construct it in the circular arc as a one-dimensional solution.

The boundary value problem \eqref{Problem_banana_1} in the conformal image $\Omega_{w}$ becomes
\beq\label{PNP_sigma_0}
\tilde v''-\frac{4\sin(\theta)(1-\cos(\theta))}{|e^{i\theta}-1-e^{i\theta}\sqrt{\eps}|^4}\tilde v'&=&-\frac{4\eps}{|e^{i\theta}-1-e^{i\theta}\sqrt{\eps}|^4}\exp \left\{-\ds \tilde v(e^{i\theta})
\right\}\label{reduced_1}\\
\tilde v'(c\sqrt{\eps})&=&\,0\label{req}\nonumber\\
\tilde v'(\pi)& =&\,-\frac{\sigma \sqrt{\eps}}{2 }.\nonumber
\eeq
To construct an asymptotic approximation to the solution of \eqref{reduced_1} in the limits $\eps\to0$ and $\sigma\to\infty$, we first construct the outer-solution in the form of a series in powers of $\eps$, which is an approximation valid away from the boundary $\theta=c\sqrt{\eps}$.  After dropping the terms in $\eps$ in \ref{PNP_sigma_0}, we obtain the outer solution by  a direct integration
\beq\label{outer_sol}
v_1(\theta)&=& -A(\theta-\sin(\theta))+\tilde v(0),
\eeq
where $\tilde v(0)$ and  $A$ are constants. The outer solution \eqref{outer_sol} cannot satisfy all boundary conditions, consequently a boundary layer correction is needed at $\theta=\pi$. An approximation of the solution can be obtained by freezing the power-law term  and neglecting the first order derivatives in \eqref{PNP_sigma_0}, for
which the equation is for a generic parameter $b>0$,
\begin{align*}
\frac{d^2 }{d \theta^2} v_b(\theta) +b  e^{\ds-v_b(\theta)}=0,\quad\frac{d v_b(0)}{d \theta}= v_b(0)=0.
\end{align*}
The solution is \cite{PhysD2016}
\beq \label{blowups}
v_b(\theta)&=&\ln\cos^2\left(\frac{b}{2}\,\theta\right).
\eeq
Putting the outer and boundary layer solutions together gives the uniform asymptotic approximation
\beq\label{unif}
y_{\mbox{\scriptsize unif}}(\theta)&=&-A(\theta-\sin(\theta))+\tilde v(0)+\ln\cos^2\left(\frac{b}{2}\,\theta\right).
\eeq
The condition at {$\theta=\pi$} gives that
\beq\label{cond}
{y_{\mbox{\scriptsize unif}}'(\pi)=-2A -b\tan\frac{b}{2}\pi= \ds-\frac{\sigma_\eps
		\sqrt{\eps}}{2}.\nonumber}
\eeq
The compatibility condition for  \eqref{NewPb},
\beq\label{Compatibility2new}
\lambda_{\eps} + \lambda_{bulk}&=& \int\limits_{\Omega}\exp \{-  u(\x)\} dS_{\x},
\eeq
gives in {$\Omega_w$} that
\beq\label{Compatibility2}
\lambda_{\eps} + \lambda_{bulk}&=& \int\limits_{{\Omega_w}}\exp \{-\tilde v(w)\} \frac{dw}{|\phi'(\phi^{-1}(w))|}.
\eeq
Using the uniform approximation \eqref{unif} in the compatibility condition \eqref{Compatibility2}, we obtain the second condition
\beq \label{Compatibility3}
\lambda_{\eps} + \lambda_{bulk} &=&\, 8\sqrt{\eps}\,e^{\ds-\tilde v(0)}\int\limits_{c\sqrt{\eps}}^\pi \frac{1}{\cos^2\ds\frac{b}{2}\theta}\frac{{\exp\left\{\ds
		A(\theta-\sin(\theta))\right\}}}{| e^{i\theta}(1-\sqrt{\eps}) - 1  |^4 }\,
d\theta\nonumber \\
&\approx&\, \frac{8\,e^{\ds-\tilde v(0)}}{\eps}\int\limits_{0}^{\pi/\sqrt{\eps}} \frac{1}{\cos^2\ds\frac{b}{2}\sqrt{\eps}\xi}\frac{
	\exp\{  A(\sqrt{\eps}\xi-\sin(\sqrt{\eps}\xi))\}}{|1+\xi^2  |^2 }\,d\xi,
\eeq
where we used the change of variable $\theta=\sqrt{\eps}\xi$. Integrating by parts, we get for $\eps\ll1${\small
\beq\label{Compatibility3b}
\lambda_{\eps} + \lambda_{bulk} &\sim& \frac{8\,e^{\ds-\tilde v(0)}}{\eps} \left( \frac{2}{b\sqrt{\eps}}\tan\frac{b}{2}\pi\frac{ e^{\ds A\pi}}{\left|1+\left(\ds\frac{\pi}{\sqrt{\eps}}\right)^2
	\right|^2 }-\int\limits_{0}^{\pi/\sqrt{\eps}} \frac{2}{b\sqrt{\eps}}\tan\frac{b}{2}\theta\ {\Psi(\theta)\,d\theta}\right),
\eeq}
where
\beq
\Psi(\xi)={
	\frac{d}{d\xi}\frac{ \exp\{  A(\sqrt{\eps}\xi-\sin(\sqrt{\eps}\xi))\}}{|1+\xi^2  |^2 }}.
\eeq
Thus, it remains to solve the asymptotic equation
\beq\label{Compatibility3b1}
\lambda_{\eps} + \lambda_{bulk} \sim 8 \,e^{\ds-\tilde v(0)}\eps^{1/2} \left[ \frac{2}{b\pi^4}\tan\frac{\pi  b}{2}\exp\{A\pi\}+ O\left(\ln\left|{\cos\ds\frac{\pi b}{2}}\right|\right)\right].
\eeq
for $A$ and $b$ in the limit $\eps\to0$.
We consider the limiting case where
\beq\label{Case_1}
\frac{A}{\sigma_{\eps}\sqrt{\eps}} \ll{1\hspace{0.5em}\mbox{for}\ \sigma_{\eps}\to\infty,}
\eeq
for which condition {\eqref{cond}} can be simplified and gives to leading order
\beq
b\tan{\frac{\pi b}{2}}=\frac{\sigma_{\eps} \sqrt{\eps}}{2 }\label{btan},
\eeq
that is, for $\sigma_{\eps}\sqrt{\eps}\ll1$ \eqref{btan} gives
$$b\approx 1-\frac{4}{\pi} \frac{1}{\sigma_{\eps}\sqrt{\eps}},\quad
\tan{\frac{b}{2}\pi\sim} \frac{\sigma_{\eps}\sqrt{\eps}}{2 }.$$
It follows from \eqref{Compatibility3b1} using \eqref{sigma_notation} that
\beq
A&=&-\frac{1}{\pi}\left(\ln\left( \frac{8\eps }{\pi^4|\p\Omega_{\eps}|\left(1 +\ds \frac{\lambda_{bulk}}{\lambda_{\eps}}\right)} \right) -\tilde v(0)\right).
\eeq
We conclude from expression \ref{unif} that
{\small
	\beq\label{unif_uncharged}
	y_{\mbox{\scriptsize unif}}(\theta)&=& \frac{1}{\pi}\left(\ln\left( \frac{8\eps }{\pi^4|\p\Omega_{\eps}|\left(1 +\ds \frac{\lambda_{bulk}}{\lambda_{\eps}}\right)} \right) -\tilde v(0)\right)(\theta-\sin(\theta))\\
	&&+\ln\cos^2\left(\ds\frac{1-\ds\frac{4}{\pi} \frac{1}{\sigma_{\eps}\sqrt{\eps}}}{2}\,\theta\right)+\tilde v(0)\nonumber.
\eeq
}We compare in Fig. \ref{f:uncharged}A-D, the uniform approximation \eqref{unif_uncharged} with numerical simulations of the reduced eq. \eqref{PNP_sigma_0} and the three-dimensional numerical solution (eq\ref{NewPb}). The difference of potential $\tilde \Delta y_{\mbox{\scriptsize unif}}=y_{\mbox{\scriptsize unif}}(0)-y_{\mbox{\scriptsize unif}}(\pi)$, can now be  estimated using \eqref{unif_uncharged} and we obtain
\beq\label{Diff_pot_uncharged}
\tilde \Delta y_{\mbox{\scriptsize unif}}&=&  -\left(\ln\left(\frac{8\eps }{\pi^4|\p\Omega_{\eps}|\left(1 +\ds \frac{\lambda_{bulk}}{\lambda_{\eps}}\right)} \right) -\tilde v(0)\right) -\ln\sin^2\left(\ds\ds \frac{2}{\sigma_{\eps}\sqrt{\eps}}\right) .
\eeq
In the small $\eps$ limit, the constant $ \tilde v(0)=O(1)$ can be neglected. We compare the analytical expression for difference of potential \eqref{Diff_pot_uncharged} with the result of the reduced equation \eqref{PNP_sigma_0} computed numerically in Fig. \ref{f:uncharged}E. We note that the solution in 3D differs from 2D, as shown in Fig. \ref{f:uncharged}F.
\begin{figure}[http!]
	\center
	\includegraphics[scale=0.82]{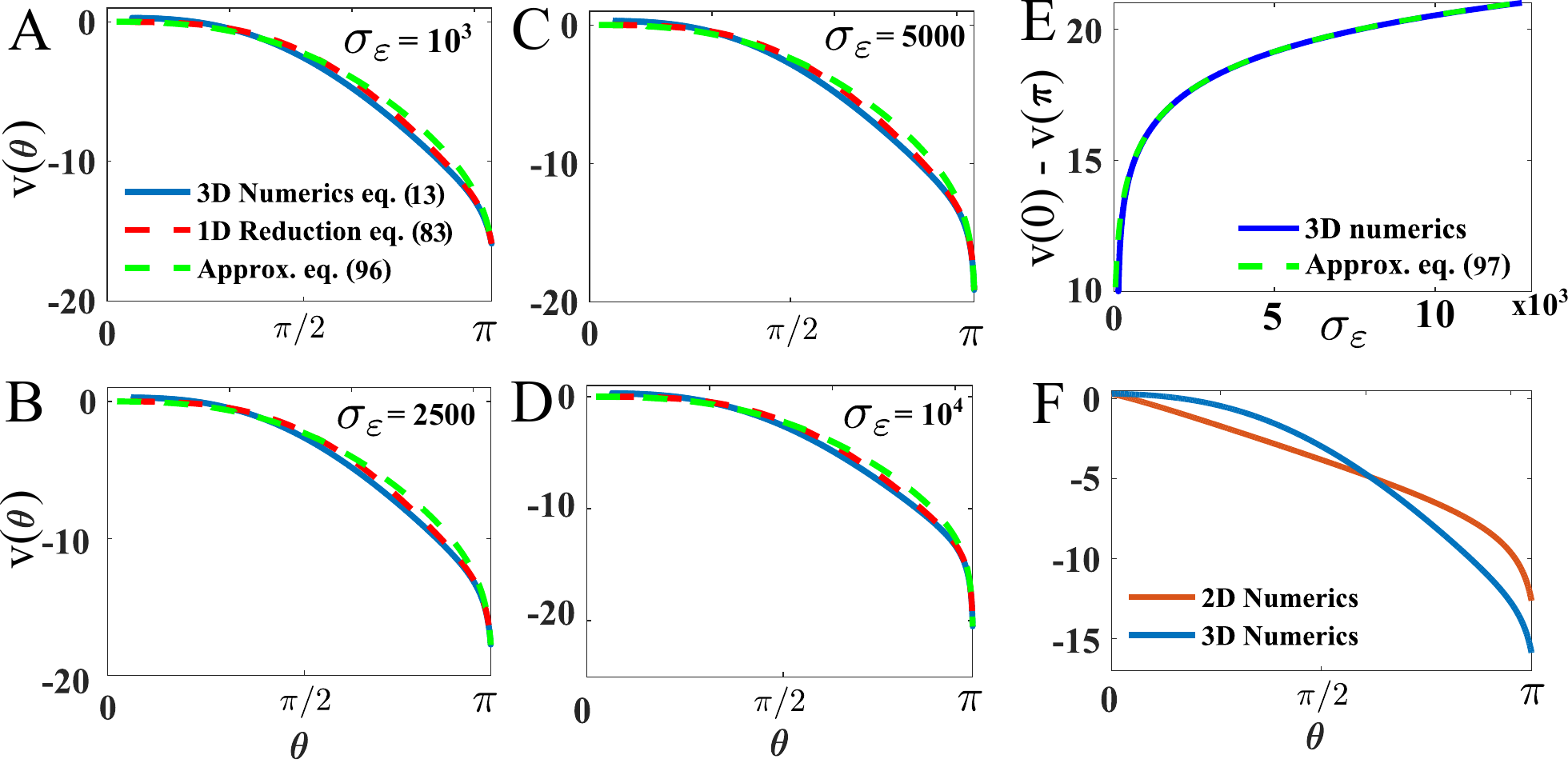}
	\caption{ {\bf \small Numerical \eqref{NewPb}-\eqref{PNP_sigma_0} versus analytical \eqref{unif_uncharged} solutions with zero Neumann boundary conditions, except at the end of the funnel. }
	 {\bf A-D} Analytical (dashed green) obtained from \eqref{unif_uncharged} and numerical solutions \eqref{NewPb} (blue) computed in 3D and the 1D reduced equation \eqref{PNP_sigma_0} (dashed red).
	 {\bf E.}  Potential difference $v(0)-v(\pi)$ computed numerically from \eqref{PNP_sigma_0} (blue) and the asymptotics \eqref{Diff_pot_uncharged}.
	 {\bf F. } Comparison of eq. \eqref{NewPb} numerical solutions in 2D (red) and 3D (blue).\label{f:uncharged}}
\end{figure}

\section{PNP solution for $\sigma_{\eps} \neq \sigma_{cusp}$ and $\sigma_{bulk}=O(1)$  }\label{s:sigma_diff}
\subsection{Analytical representation of the PNP solution}
We study here the effect of the charge density $\sigma_{cusp}$ located on the cusp-shaped funnel on the solution $u(\x)$ of
\beq\label{NewPb2}
-\Delta {u}(\x)&=&\,\exp\{-{u}(\x)\}\hspace{0.5em}\mbox{for}\ \x\in \Omega\\
 \frac{\p u(\x)}{\p n}&=&-\sigma_{\eps}  \mbox{ on } \p\Omega_{\eps}\nonumber\\
 \frac{\p u(\x)}{\p n}&=& -\sigma_{cusp}  \mbox{ on } \p\Omega_{cusp} \nonumber
 \\
 \frac{\p u(\x)}{\p n}&=&  -\sigma_{bulk} \mbox{ on } \p\Omega_{bulk},\nonumber
\eeq
in the small $\eps$ and large $\sigma_{cusp}$ limits, such as $\sigma_{cusp}\sqrt{\eps}\gg1$, $\sigma_{\eps}/\sigma_{cusp}=O(1)$ and $\sigma_{bulk}=O(1)$.

In the large $\sigma_{cusp}\sqrt{\eps}$ limit, we have shown (section \ref{s:2}) that for $\theta$ in the range $[c\sqrt{\eps}, \pi-\sqrt{\eps}]$ (region $A$, Fig. \ref{f:RegionsAB}), the angular derivatives of
$u_{unif}$ can be neglected.  We thus use the result of eq. \eqref{Solution_uA_theta} by changing $\sigma$ by $\sigma_{cusp}$ to obtain
\beq\label{Solution_In_Acusp}
u_{cusp}(  \rho,\theta )&=& \ln\left( 2\left(\frac{2 (1-\cos(\theta )) +\sigma_{cusp}\eps}{\pi\sigma_{cusp}\eps}\right)^2\right)+\ln\left(\frac{4\eps^2 }{|e^{i\theta }-1|^4}\right)\\
&&+\ln\left(\cos^2\left(    \frac{\pi\sigma_{cusp}(\rho -\eps)}{2(2 |(1-\cos(\theta ))+\sigma_{cusp}\eps)}\right)\right)\nonumber \\
&& +O(\sqrt{\eps}).\nonumber
\eeq
For $\rho=1-\sqrt{\eps}$ and $\theta\in[c\sqrt{\eps}, \pi-\sqrt{\eps}]$, we get
\beq\label{uA_new_sigma}
u_{cusp}( 1-\sqrt{\eps},\theta )&=& \ln\left( 8\left(\frac{2 (1-\cos(\theta )) +\sigma_{cusp}\eps}{\pi\sigma_{cusp}|e^{i\theta }-1|^2}\right)^2\right)+O(\sqrt{\eps}).
\eeq
To construct a uniform solution $u_{unif}$, we match to a solution $u_B$ in region\\ $B=\{(\rho,\theta), \theta\in[ \pi-\sqrt{\eps},\pi] \hbox{ and } \rho=1-\sqrt{\eps} \}$. We obtain the general expression
\beq\label{Sol_unif_sigma_diff}
u_{unif}(\rho,\theta)&=&
\begin{cases}\label{diff_cusp_eps}
	 \ln\left( \ds8\left(\frac{2 (1-\cos(\theta )) +\sigma_{cusp}\eps}{\pi\sigma_{cusp}|e^{i\theta }-1|^2}\right)^2\right)  &\hbox{ for }  \theta \in [0, \pi-\sqrt{\eps}] \\
	\\
	u_{B}(\theta) &\hbox{ for }  \theta \in [\pi-\sqrt{\eps},\pi].
\end{cases}
\eeq
Thus the difference of potential $u(C)-u(S)$ between the center of mass $C$ and the funnel base $S$ is then
\beq\label{Diff_sigma_diff}
V(C)-V(S)&=& -\ln\frac{2\sigma_{cusp}\pi^2}{(4 +\sigma_{cusp}\eps)^2}+\ln\frac{\pi^2c^4}{8} + \tilde\Delta u_B +O\left(\eps,\frac{1}{\sigma_{cusp}}\right),
\eeq
where $c\sqrt{\eps}$ is the angular coordinate of the mapped center of mass $C$ in $\Omega_w$. We compare in Fig. \ref{f:Surface_part}A-B the analytical expression (dashed) of \eqref{Sol_unif_sigma_diff} with the three-dimensional numerical simulations (solid) of $u$ (eq. \eqref{NewPb2}). When $u_B$ is given by expression \eqref{uB_C0} with condition $\sigma_{\eps}\sqrt{\eps}\gg1$, then the difference of potential is given by
\beq\label{Delta_diff_ub_large}
u(C)-u(S)&=&\ds\frac{k T}{e}\left(\ln\sin^2\frac{\pi  }{ \sigma_{ \eps }\eps}-\ln\frac{2\sigma_{cusp}\pi^2}{(4 +\sigma_{cusp}\eps)^2}\right)+O(1).
\eeq
The two conditions $\sigma_{\eps}/\sigma_{cusp}=O(1)$ and $\sigma_{bulk}=O(1)$ imply that the uniform solution is not affected by the bulk or the tip of the cusp. This is in contrast with the results computed for $\sigma_{cusp}=0$ (section \ref{s:Null}) for which the solution in the cusp is entirely defined by the surface charge densities $\sigma_{cusp}$ and $\sigma_{bulk}$ (see eq. \eqref{unif_uncharged}). However, when the previous conditions are not satisfied ($\sigma_{\eps}/\sigma_{cusp}=O(1)$ is not verified), the numerical solution (red) and the analytical expression \eqref{Sol_unif_sigma_diff} (dashed blue) do not agree (Fig. \ref{f:Surface_part}A-B).
\subsection{PNP solution with reflecting boundary at the end of the funnel}
When we impose a reflecting boundary condition at the end of the cusp $\p\Omega_{\eps}$ ($\sigma_{\eps}=0$),
we construct an approximation of equation \eqref{NewPb2} in the regimes $\eps\ll1$ and $\sigma_{cusp}\gg1$  in the following regime of parameters $\sigma_{cusp}\sqrt{\eps}\gg1$ and $\sigma_{bulk}=O(1)$.\\
To construct the approximation $u_{unif}$ in $\Omega_w$, we use expression in the cusp \eqref{Sol_unif_sigma_diff}, where the solution $u_B$ is constructed by extending $u_{cusp}( \rho,\theta )$ to region $B$. We have
\beq\label{u_A_extenddiff}
\left.\frac{\p u_{unif}(\rho, \theta)}{\p \theta}\right|_{\theta=\pi} &=&0.
\eeq
To show that $u_{cusp}$  satisfies the same boundary condition, we differentiate $u_{cusp}( \rho,\theta )$, (eq. \eqref{Solution_In_Acusp}), in $\theta$ at $\theta=\pi$:
\beq\label{u_A_extend}
\left.\frac{\p u_{cusp}(\rho, \theta)}{\p \theta}\right|_{\theta=\pi} &=&0.
\eeq
We conclude that $u_{cusp}$ matches at $\theta=\pi$ the boundary condition satisfied by the solution $u(\x)$ for $\sigma_{\eps}=0$. Consequently,
\beq\label{Sol_unif_sigma_diff2}
u_{unif}(\rho,\theta)&=&
	\ln\left( \ds8\left(\frac{2 (1-\cos(\theta)) +\sigma_{cusp}\eps}{\pi\sigma_{cusp}|e^{i\theta}-1|^2}\right)^2\right).
\eeq
Thus the difference of potential between the funnel base $S$ and the center of mass $C$ is
\beq\label{Diff_sigma_diff2}
u(C)-u(S)&=& -\ln\frac{2\sigma_{cusp}a^2\pi^2}{(4 +\sigma_{cusp}\eps)^2}+\ln\frac{\pi^2c^4}{8}.
\eeq
We obtain a good agreement between the analytical expression (eq. \eqref{Sol_unif_sigma_diff2}) and
 the three dimensional numerical solution of \eqref{NewPb2} (Fig. \ref{f:Surface_part}C).

The result obtained from \eqref{Diff_sigma_diff2} can be used to model the voltage in a domain with a cusp-shaped funnel connecting a reservoir with a fixed electrical potential and zero electric field at the of funnel-reservoir junction. This no field condition is satisfied when $\sigma_{ \eps}=0$. This result can be applied to the electrical properties of dendritic spines with a short neck (see \cite{YusteBook}, p.28, Fig. 3.9, spine 7), approximated by a cusp and the parent dendrite as a reservoir.
\begin{figure}[H]
	\center
	\includegraphics[scale=0.75]{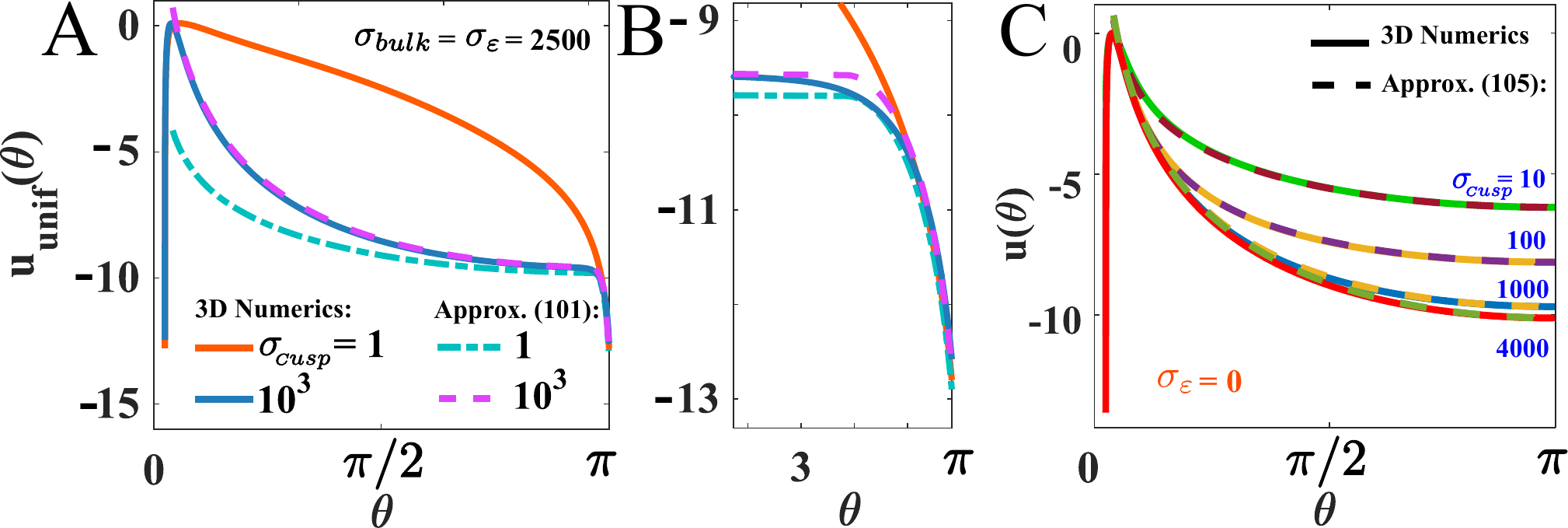}
	\caption{ {\small  {\bf Comparison of numerical and analytical solutions for non-homogeneous surface charge density}
	 {\bf A. } Numerical (eq. \eqref{NewPb2}) in 3D (solid) and analytical (eq. \eqref{Sol_unif_sigma_diff})  (dashed) solutions for $\sigma_{cusp}=1000$ and $\sigma_{cusp}=1$, $\sigma_{bulk}=\sigma_{\eps}=2500$.
	 {\bf B. } Magnification of panel A in the region of $\theta=\pi$.
	 {\bf C. } 3D Numerical (solid) from eq. \eqref{NewPb2} and analytical (eq.\eqref{Sol_unif_sigma_diff2}) (dashed) solutions computed for $\sigma_{cusp}=10,\,100, \,1000$ and $4000$, where $\sigma_{\eps}=0$ and $\sigma_{cusp}= \sigma_{bulk}$. Here $\eps=0.01$.
  }\label{f:Surface_part}}
\end{figure}
\section{Discussion and conclusion}
Based on the steady-state solution of the Poisson-Nernst-Planck equations, we derived here electrostatic properties of non-electro-neutral electrolytes confined in a cusp-shaped funnel geometry. We showed that the local curvature and the distribution of surface charges shape the electrical landscape within small domains. The new electrical properties have been obtained for a dominant ionic specie, in an electrolyte having an excess of charges in two dimensions in \cite{NonLin2017}. The new mathematical methods consist here in the construction of an asymptotic expansion of the nonlinear PNP equations inside 3D domains, with non-homogeneous Neumann boundary conditions.

Using asymptotics methods validated by numerical solutions of the PNP equations, we found several explicit voltage drops: first, for a surface charge density homogeneously distributed, the electrical potential distribution in 3D and 2D domains is quite similar to leading order potential inside a planar cusp (Fig. \ref{f:3D_Cusp}). However, the voltage inside an uncharged funnel (Fig. \ref{f:uncharged}), associated to the condition $\sigma_{cusp}=0$ varies significantly between a 2D and 3D domain with a cusp funnel.
We summarize in table \ref{t:sol} the results we have obtained in the three sections above, where we reintroduce the physical units and used $\sigma_i=\tilde \sigma_i z e R_c/kT$ (section \ref{s:2}).

The presence of negative ions in $\Omega$ may slightly reduces the voltage. However, as shown in \cite{NonLin2017}, accounting for negative charges carried by chloride anions present in the cytosol at physiological concentration \cite{Hille}, does not alter the voltage to leading order. Consequently the voltages summarized in table \ref{t:sol} provide insights for understanding the electro-diffusion properties. The present results could be used in the design accurately quartz nanopipettes with an optimal shape \cite{HS2012,Perry,Jayant}. It would be interesting to vary the surface charge densities \cite{Sparreboom} in some sub-regions $\sigma_{bulk}$, $\sigma_{cusp}$ or $\sigma_{ \eps}$ \eqref{2_BL_conditions}.

Finally, the present analytical results can be used to predict the voltage drop in neuronal microdomains such as dendritic spines \cite{YusteBook}.  The local curvature is certainly a  key factor in modulating the voltage and thus we are beginning to understand how nano- and micrometer geometry can encode synaptic modulation, that underlyes learning and memory in the brain. Indeed, in compartment such as dendritic spines, the high curvature variation play a major role in converting injected current into voltage. This effect may as well influence the propagation and genesis of local depolarization in excitable cells \cite{HY2015,Rall,Qian}.
\begin{table}[H]
\begin{center}
\begin{tabular}{|c|c|  }
	\hline
	Conditions & $  V(C)-V(S)$   \\
	\hline
	& \\
	 $\tilde\sigma\gg1$ & $\ds\frac{k T}{e}\left (\ln\sin^2\frac{kT\pi}{e \tilde\eps\tilde \sigma}-2\ln\frac{\sqrt{2}\,e\pi
     R_c\tilde{\sigma} }{   4kT  +  e\tilde \eps\tilde {\sigma}   }+O(1) \right)$ \\
	& \\
	
	 $\begin{matrix}
	\tilde\sigma_{cusp}=0\\
	\tilde\sigma_{\eps}\gg1\\
	\tilde\sigma_{bulk}=C^{ste}
	\end{matrix}$& $\ds \frac{k T}{e}\left (- \ln  \frac{8 R_c\tilde\eps}{ \pi^4|\p\tilde\Omega_{\eps}|\left(1 +  \tilde{\sigma}_{bulk}/ \tilde{\sigma}_{\eps}\right)}    -\ln\sin^2 \ds \frac{2 kT}{e\tilde{\sigma}_\eps\sqrt{ R_c \tilde\eps}} +O(1) \right)$ \\
	& \\
	
	$\begin{matrix}
	\tilde\sigma_{cusp}\neq \tilde\sigma_{\eps}\\
	\tilde\sigma_{cusp}\sqrt{\eps}\gg1\\
	\tilde\sigma_{bulk}=C^{ste}
	\end{matrix}$ & $\ds\frac{k T}{e}\left(\ln\sin^2\frac{kT\pi}{e \tilde\eps\tilde \sigma_{\eps}}-2\ln\frac{\sqrt{2}\,e\pi
		R_c\tilde{\sigma}_{cusp} }{   4kT  +  e\tilde \eps\tilde {\sigma}_{cusp}   }+O(1)  \right)$   \\
	 &  \\
	
	
	\hline
\end{tabular}
\caption{Electrodiffusion laws for voltage drop for various surface charge densities\label{t:sol}}
\end{center}
\end{table}
\section{Appendix}\label{Appendix1}
\subsection{Radial derivative under the Mobius map \eqref{Mobius}}
We shall describe in this appendix the computation step to reduce the first order radial derivative from \eqref{Problem_Cusp2} leading to the result \eqref{Problem_banana_1} in section  \ref{s:Reduction_cplx}. First, we note that in complex coordinates, we have
\beq\label{diff_r0}
\frac{\p u( \tilde r, \tilde z)}{\p \tilde r} &=& \Re e \left(  \nabla u( \xi)\right),
\eeq
where we define
\beq\label{gradient0}
\nabla u(\xi)&=&\frac{\p u( \tilde r, \tilde z)}{\p  \tilde r} +i\frac{\p u( \tilde r, \tilde z)}{\p \tilde z}.
\eeq
Under the conformal mapping \eqref{Mobius}, the gradient \eqref{gradient0} is transformed as follows
\beq\label{grad_trans0}
\nabla u(\xi) &=& \nabla_w v(w) \, \overline{w'(\xi)}.
\eeq
Using the notation $w=X+iY$, we get 
\beq\label{grad_trans20}
\nabla_w v(w)&=& \frac{\p v(X,Y) }{\p X } + i \frac{\p v(X,Y)}{\p Y }.
\eeq
We define the real functions $w_1(X,Y)$ and $w_2(X,Y)$ that satisfy
\beq\label{w1_w20}
\overline{w'(\xi)}&=\overline{w'(w^{-1}(X,Y))}&=w_1(X,Y)+i\,w_2(X,Y).
\eeq
Using \eqref{Mobius} (M\"obius transformation), we get
\beq\label{w1_w2_construct0}
\overline{w'(w^{-1}(X,Y))}&=&\frac{\overline{(1+\alpha w)^2}}{1-\alpha}.
\eeq
Equations \eqref{w1_w20} and \eqref{w1_w2_construct0} lead to
\beq
w_1(X,Y) &=& \frac{(1+\alpha X)^2-\alpha^2Y^2}{1-\alpha^2}\\
w_2(X,Y) &=&-\frac{2\alpha Y(1+\alpha X)}{1-\alpha^2}. \nonumber
\eeq
From \eqref{diff_r0}-\eqref{grad_trans0}-\eqref{grad_trans20}-\eqref{w1_w20}, we obtain
\beq\label{diff_r20}
\frac{\p u}{\p \tilde r} &=& w_1(X,Y)\frac{\p v(X,Y) }{\p X }-w_2(X,Y)\frac{\p v(X,Y) }{\p Y }.
\eeq
Due to the round geometry of the banana-shaped domain $\Omega_w$, it is convenient to switch from Cartesian coordinates $(X,Y)$ to polar coordinates $(\rho, \theta)$. Setting $v(X,Y)=\tilde v(\rho, \theta)$, we get
\beq\label{diff_decomp0}
\frac{\p v(X,Y) }{\p X } &=& \frac{\p \tilde v(\rho,\theta) }{\p \rho }\frac{\p \rho }{\p X }+\frac{\p \tilde v(\rho,\theta) }{\p\theta }\frac{\p \theta }{\p X }\\
\nonumber \\
\frac{\p v(X,Y) }{\p Y } &=&\frac{\p \tilde v(\rho,\theta) }{\p \rho }\frac{\p \rho }{\p Y }+\frac{\p \tilde v(\rho,\theta) }{\p\theta }\frac{\p \theta }{\p Y }  \nonumber
\eeq
where,
\beq\label{polar_trans0}
\frac{\p \rho }{\p X } = \cos(\theta) & \mbox{  ,  }&\frac{\p \rho }{\p Y } =\sin(\theta) \\
\frac{\p \theta }{\p X } = -\frac{\sin(\theta)}{\rho} & \mbox{  ,  }&\frac{\p  \theta }{\p Y } = \frac{\cos(\theta)}{\rho}. \nonumber
\eeq
Using \eqref{diff_decomp0} and \eqref{polar_trans0} in \eqref{diff_r20}, it follows that
\beq\label{diff_r30}
\frac{\p \tilde u(\tilde r, \tilde z)}{\p \tilde r} &=& \frac{\p  \tilde v(\rho,\theta)}{\p \rho }\left( \cos(\theta) \tilde w_1(\rho,\theta)  - \sin(\theta)\tilde w_2(\rho,\theta)  \right) \\
&&-\frac{1}{\rho}\frac{\p  \tilde v(\rho,\theta) }{\p \theta }\left( \sin(\theta)\tilde w_1(\rho,\theta) +\cos(\theta) \tilde w_2(\rho,\theta) \right), \nonumber
\eeq
where we set $\tilde w_i(\rho,\theta)=w_i(X,Y)$ for $i\in\{1\,,\,2\}$, such as
\beq\label{w1_w2_20}
\tilde w_1(\rho,\theta)&=& \frac{1-\alpha^2 \rho^2 +2\alpha \rho \cos(\theta)(1+ \alpha  \rho   \cos(\theta)) }{1-\alpha^2}\\
\nonumber  \\
\tilde w_2(\rho,\theta)&=& -2\alpha \rho\sin(\theta)\frac{ 1 + \alpha  \rho   \cos(\theta) }{1-\alpha^2}. \nonumber
\eeq
Using \eqref{w1_w2_20} and \eqref{diff_r30}, we obtain to leading order
\beq\label{Diff_order_10}
\frac{1}{\tilde r}\frac{\p \tilde u(\tilde r,\tilde z)}{\p \tilde r}&=&
-  \frac{ \rho(1-\cos(\theta))^2}{\eps^{3/2}}\frac{\p\tilde v (\rho,\theta z)}{\p\rho}
- \frac{\sin(\theta)(1-\cos(\theta))}{\eps}\frac{\p\tilde v (\rho,\theta z)}{\p\theta}. 
\eeq
\subsection{The numerical procedure}
Numerical solutions were constructed by the COMSOL Multiphysics 5.0 (BVP problems), Maple 2015 (Shooting problems) and Matlab R2015 (Conformal mapping). The boundary value problems in 1D, 2D, and 3D were solved by the finite elements method in the COMSOL 'Mathematics' package. We used an adaptive mesh refinement to ensure numerical convergence for large value of the parameters $\sigma$, $\sigma_{\eps}$, $\sigma_{bulk}$ and $\sigma_{cusp}$.
We solved the PDEs by the shooting procedure for boundary value problems using Runge-Kutta fourth-order  method.

%
%
%
%
%
\newpage
\bibliographystyle{plain}

\end{document}